\documentclass{aa}  

\usepackage{graphicx}
\usepackage{xcolor}
\usepackage{subcaption}
\usepackage{txfonts}
\usepackage{hyperref}
\hypersetup{
    colorlinks=true,
    linkcolor=blue,
    filecolor=blue,
    urlcolor=blue,
    citecolor=blue,
}
\bibpunct{(}{)}{;}{a}{}{,} % to follow the A&A styl

\newcommand{\unit}[1]{\ensuremath{\;{\mathrm{#1}}}}

\begin{document}

   \title{Turbulence in the intragroup and circumgalactic medium}

%   \subtitle{I. Overviewing the $\kappa$-mechanism}

   \author{W. Schmidt
          \inst{1}
          \and
          J. P. Schmidt
          \inst{1}
          \and
          P. Grete\inst{2}
          }

   \institute{Hamburger Sternwarte, Universität Hamburg, Gojenbergsweg 112,
             D-21029 Hamburg, Germany\\
             \email{wolfram.schmidt@uni-hamburg.de}
         \and
             Department of Physics and Astronomy,
             Michigan State University, East Lansing, MI 48824, USA\\
             \email{grete@pa.msu.edu}
             }

   %\date{Received September 15, 1996; accepted March 16, 1997}
   \date{Preprint}

% \abstract{}{}{}{}{} 
% 5 {} token are mandatory
 
  \abstract
   {In massive objects, such as galaxy clusters, the turbulent velocity dispersion, $\sigma_\mathrm{turb}$, is tightly correlated to both the object mass, $M$, and the thermal energy.}
   {Here, we investigate whether these scaling laws extend to lower-mass objects in dark-matter filaments.}
  % methods hear there is a bhyding (mandatory)
   {We perform a cosmological zoom-in simulation of a filament using an adaptive filtering technique for the resolved velocity component and a subgrid-scale model to account for the unresolved component. We then compute the mean turbulent and thermal energies for all halos in the zoom-in region and compare different definitions of halo averages. Averaging constrained by density and temperature thresholds is favored over averages solely based on virial spheres.}
   {We find no clear trend for the turbulent velocity dispersion versus halo mass, but significant correlation and a scaling law with exponent $\alpha\sim 0.5$ between the turbulent velocity dispersion and thermal energy that agrees with a nearly constant turbulent Mach number, similar to more massive objects.}
   {We conclude that the self-similar energetics proposed for galaxy clusters extends down to the CGM of individual galaxies.}


   \maketitle
%
%________________________________________________________________

\section{Introduction}

In contrast to galaxy clusters, groups consist of a small number of galaxies and their halos have a typical mass $\sim 10^{13} M_\odot$. Most importantly, groups of galaxies along with isolated galaxies are mainly found in the dark-matter filaments of the cosmic web \citep{Lietzen2012,Cautun2014,Tempel2014}. Even so, they host nearly half of all galaxies and there are observations indicating that the intragroup medium (IGrM) constitutes a significant fraction of the baryons in the Universe \citep{Mulchaey2000,Freeland2011}. Observations of groups also suggest that they are typically out of equilibrium \citep{OSullivan2014}. In particular, scaling laws for the X-ray luminosity of groups are observed to deviate from clusters \citep{Bharadwaj2014,Vajgel2014,Lovisari2015} in two ways. First, groups exhibit a more pronounced scatter than clusters.  Second, there exist some indications of a change of slope in the luminosity-mass and luminosity-temperature relations.

In numerical simulations, various processes affecting the IGrM have been identified. For example, the rate of galaxy mergers is relatively high in groups \citep{Diaz-Gimenez2010}. Outflows produced by active galactic nuclei (AGN) and supernovae, which are enhanced by mergers, are expected to have a significant impact on thermal and non-thermal properties of IGrM \citep{Mihos1996,Planelles2013,Liang2016,Martin2018,Patton2020}. In addition to tides and outflows, galaxies interact via gas stripping with the surrounding medium \citep{Iapichino2008,Roediger2015}. All of these processes produce turbulence and heat the IGrM.

By computing luminosity relations for simulated groups, \citet{Liang2016} and \citet{Paul2017} were able to confirm deviations from the scaling that follows from self-similar structure formation. Objects in the mass range of groups show a steeper slope and relatively high entropy. This suggests that the energy budget of the IGrM is not predominately controlled by the depth of the gravitational potential well. Indeed, the analysis of \citet{Paul2017} indicates that many groups are far from virial equilibrium. If this is so, tidal tails of merging galaxies, turbulent wakes produced by galaxies falling into the potential well of a group, and outflows of galaxies should elevate the kinetic energy by stirring turbulent motions in the IGrM. Since outflows are also expected for isolated galaxies, they should produce turbulence in the gas surrounding the galaxies, i.e.\ the circumgalactic medium (CGM), see, e.g., \citet{Tumlinson2017,Lochhaas2020}. 

As a result, the turbulent velocity dispersion is an important indicator of non-equilibrium conditions. Unfortunately, the computation of the turbulent velocity dispersion is nontrivial. From a physical point of view, it is important to distinguish between bulk motions (that are induced, for instance, by gravity) and fully nonlinear, turbulent flows. For example, the accretion of gas into the potential well of a halo produces bulk flows on large scales, while eddies produced by hydrodynamic instabilities in the wake of a moving galaxy or an outflow propagating into the surrounding medium are turbulent. Separating these components is not possible with the commonly applied method to compute standard deviations of the radial velocity component in spherical shells. To circumvent this issue, \citet{Schmidt2016} computed the three-dimensional turbulent velocity dispersion $\sigma_\mathrm{turb}$ by means of an adaptive algorithm that filters out bulk flows (e.g., accretion flows). It was shown that both in the intracluster medium (ICM) and in the warm-hot intergalactic medium (WHIM), the turbulent velocity dispersion follows power laws. In particular, a strong correlation of turbulent and thermal energies was established, indicating what is called second self-similarity \citep{Miniati2015,Schmidt2017}. This relation can be expressed as $\sigma_\mathrm{turb}\propto e^{\,\alpha}$, where $e$ is the specific thermal energy and $\alpha$ the power-law exponent. For the ICM, $\alpha\approx 0.5$ was found, implying that the turbulent Mach number in halos is roughly constant. Moreover, the turbulent Mach number in halos was generally found to be close to unity. This is physically expected, as turbulence is driven by accretion shocks, mergers, stellar feedback,and active galactic nuclei. Supersonic flows generated by these processes rapidly decay  to subsonic turbulence, which in turn decays more slowly.

In light of the observational and numerical results mentioned above, it is interesting to ask whether the energetics of objects in the mass range of groups and isolated galaxies reveal fundamental differences compared to more massive clusters, in other words, whether $\alpha$ changes. In this article, we use the adaptive mesh refinement (AMR) Enzo to zoom into a filament in a cosmological volume with a spatial resolution scale of $1\,\mathrm{kpc}$ at the maximum refinement level. The subgrid physics encompasses a novel subgrid-scale (SGS) model for numerically unresolved turbulence \citep{Grete2016} in combination with standard recipes for star formation and stellar feedback in galaxies. After describing the numerical methods applied in our simulations in more detail in the following Section, we present an analysis of the numerical data in Sect.~\ref{sec:results}. We begin with a phenomenological discussion and an analysis of radial profiles for some representative halos. Then we present statistics for all halos in the selected filament including an analysis of how to constrain the IGrM or, equivalently, the circumgalactic medium of isolated galaxies in simulations. Finally, we investigate the correlations of the mean turbulent velocity dispersion with the halo mass and the thermal energy. In our conclusions in Sect.~\ref{sec:conclusion}, we discuss our results with regard to the question raised above.

%________________________________________________________________

\section{Numerical methods}
\label{sec:numerics}

We use a modified version of the publicly available, open source cosmological AMR code Enzo \citep{Enzo2014,Enzo2019}.\footnote{See also website \url{enzo-project.org}. Our modifications are available at \url{https://github.com/pgrete/enzo-dev}. Commit \texttt{6300b03} was used in this work.} 
The code is MPI parallelized, features N-body dynamics based on a second-order drift-kick-drift algorithm in combination with cloud-in-cell interpolation to compute the joint gravitational potential of gas and particles, and a variety of different finite volume solvers for gas dynamics and magnetic fields. In our simulations, we apply the monotonic upstream-centered scheme for conservation laws (MUSCL) with a local Lax-Friedrichs (LLF) solver for sufficient numerical robustness. In addition, we utilize a structural subgrid-scale (SGS) model that computes effects of numerically unresolved turbulence on the basis of the instantaneous flow structure at the smallest resolved scales, see \citet{Grete2016,Grete2017} for idealized setups and \citet{Grete2019} for a cosmological application. This approach avoids the difficulties and computational cost of solving an additional partial differential equation \citep{Maier2009}. While the impact on gas properties in halos is typically negligible compared to the statistical variation among comparable objects evolving from different initial conditions~\citep{Grete2019}, it was found that the employed model improves higher-order statistics of quantities such as the vorticity in compressible turbulence simulations \citep{Grete2017}. Since turbulence is confined to the cosmic web \citep{Schmidt2016}, we implemented the Kalman filtering technique for spatially inhomogeneous turbulence into Enzo. Since no no fixed smoothing scale is applied, this method is particularly suitable for spatially inhomogeneous turbulence. In the statistically stationary regime, the filter operates like an exponential low-pass filter with a characteristic time-scale of $5\,$Gyr and a characteristic velocity scale of $100\,{\mathrm{km/s}}$ (see statistics discussed in Sect.~\ref{sec:results_halos}). As a result, Kalman filtering allows us to separate the turbulent fluctuation $\vec{v}^\prime$ from non-turbulent bulk flows such as gas accretion into the potential wells of halos and filaments: 
\begin{equation}
    \vec{v}^\prime=\vec{v}-[\vec{v}],
\end{equation}
where $\vec{v}$ and $[\vec{v}]$ is the unfiltered and filtered velocity, respectively. In combination with the SGS model for the numerically unresolved specific kinetic energy $E_\mathrm{sgs}$, we define the turbulent velocity dispersion by
\begin{equation}
  \label{eq:sigma_turb}
   \sigma_{\rm turb}^2 = |\vec{v}^\prime|^2 + 2E_\mathrm{sgs},
\end{equation}
A detailed description and numerical tests of the algorithm can be found in \citet{Schmidt2014}.

\begin{figure}
   \centering
   \includegraphics[width=0.5\textwidth]{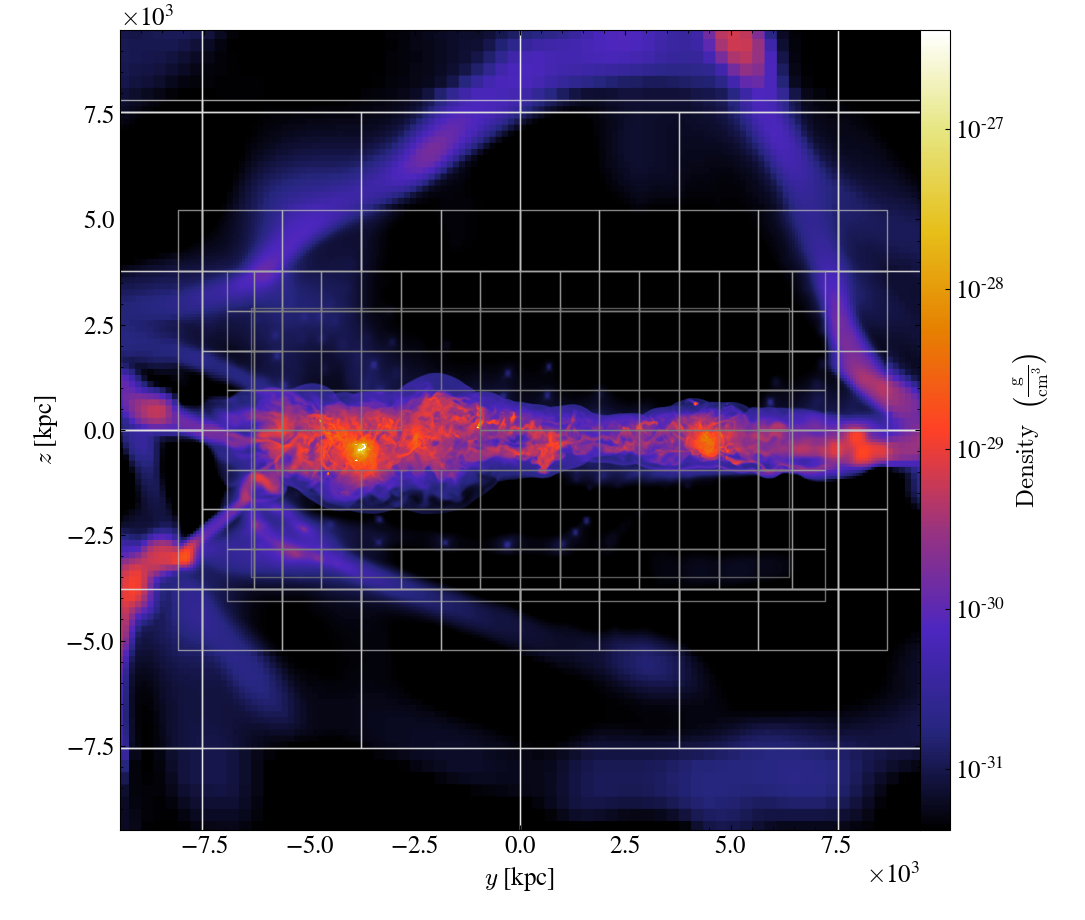}
   \caption{Slice of the gas density in a zoom-in simulation of a filament at redshift $z=0$. The shown region is a close-up view of the central quarter of the simulation box ($74.3\,\mathrm{Mpc}$ physical size). Gray lines show edges of Enzo subgrids (i.e.\ pieces distributed among processors) up to level 4. AMR levels 5-8 are hidden. }
   \label{fig:slices_filament}
\end{figure}

To achieve sufficient mass resolution for groups of galaxies and individual galaxies in a filament, the zoom-in technique is applied. With the help of MUSIC \citep{Hahn2011}, we generated initial data for a $256^3$ cosmological box of co-moving size $50\,\mathrm{Mpc}/h$ and four nested grids (levels 1-4) in a volume of about $3.9\times 8.6\times 3.9\,(\mathrm{Mpc}/h)^3$. As cosmological parameters, we chose $h=0.673$, $\Omega_{\rm b}=0.0487$, and $\Omega_{\rm m}=0.315$ from the \citet{Planck14}.\footnote{Corrections in later releases of the Planck data are not significant for this work.} The particle mass is about $8\times 10^{8}\,M_\odot$ at the coarsest (root-grid) resolution and $2\times 10^{5}\,M_\odot$ in the zoom-in region. To reach the targeted spatial resolution of $1\,\mathrm{kpc}$, up to four AMR levels (levels 5 to 8) were added in the course of the simulation by applying refinement by dark matter and baryon mass. The total number of cells in our highest-resolution run culminated above $5\times 10^8$. As an illustration, Figure~\ref{fig:slices_filament} shows a slice through the nested-grid region around a prominent filament. We use the Python package yt \citep{Turk2011} for postprocessing and visualization.

\begin{figure}
   \centering
   \includegraphics[width=0.5\textwidth]{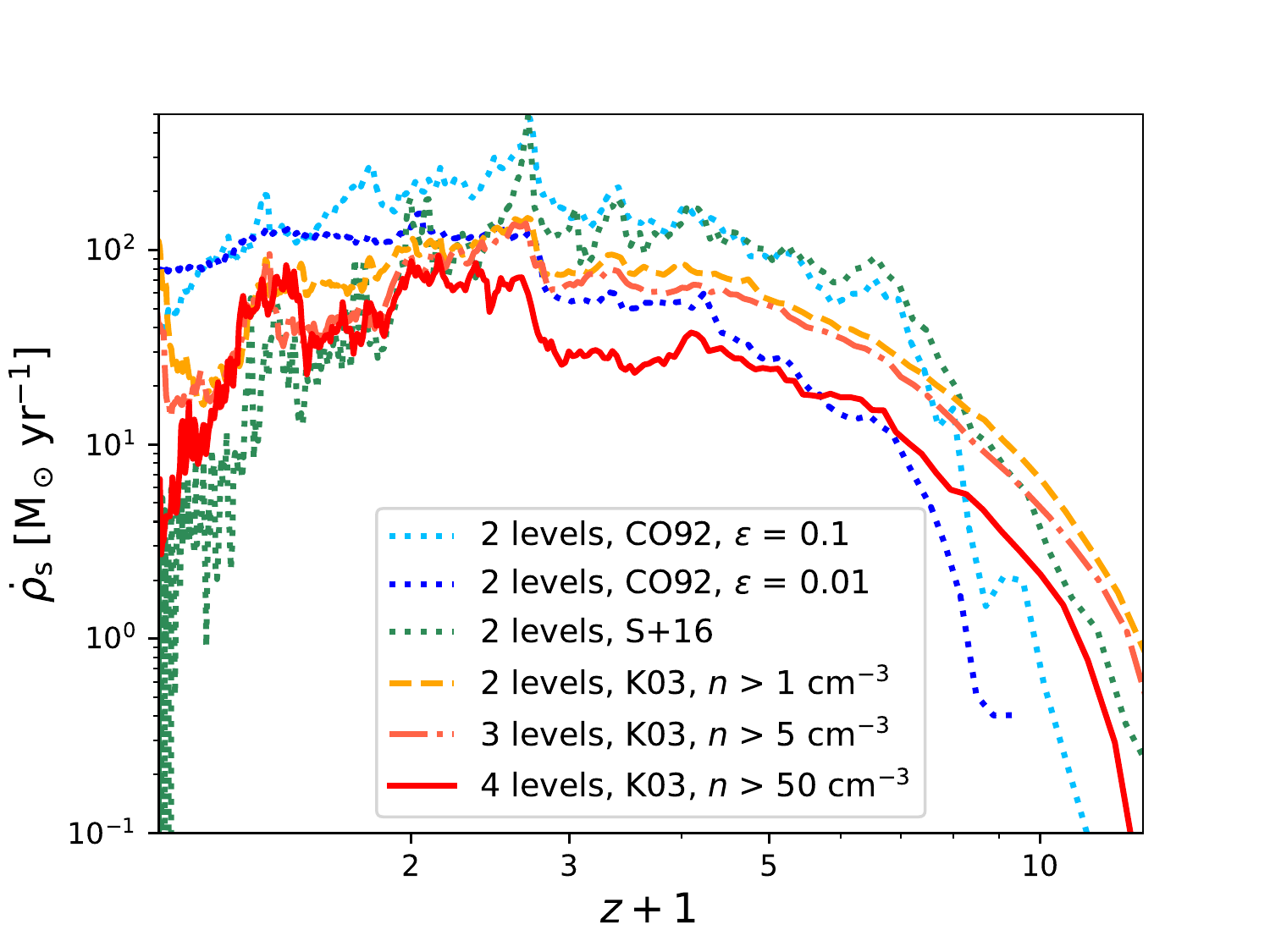}
   \caption{Mean star formation rate per unit volume in zoom-in region for different models and resolutions: \citet{Cen1992} model (CO92) with stochastic star formation, where $\varepsilon$ is the star formation efficiency, a dynamical model with local star formation efficiency (\citealt{Semenov16}, S+16), and the \citet{Kravtsov2003} model (K03) with threshold number density $n$ and a fixed star formation time scale of $1\,\mathrm{Gyr}$. The number of levels specified in the legend refers to the number of AMR levels on top of nested-grid levels. Our fiducial model is shown as solid (online version: red) line.}
   \label{fig:sfr}
\end{figure}

To treat chemical species and radiative cooling, we use the Grackle library. In our simulations, we applied a 6-species atomic H and He network, metal cooling using the Cloudy tables, and heating and cooling rates and UV background rates from \citet{Haardt2012}.\footnote{For further details, see \url{grackle.readthedocs.io/en/latest/Parameters.html}} In addition to cooling, star formation and feedback are crucial ingredients to determine the thermodynamic state of the gas. Unfortunately, star formation is notoriously difficult to model in cosmological simulations \citep[see review by][]{Naab2017}.  For this reason, we evaluated three models in test runs at lower resolution (two AMR levels, six levels of refinement in total) in preparation for our fiducial run. 
\footnote{Usually, the total star formation rate in a cosmological volume with periodic boundary conditions is normalized to stellar mass per unit time and unit volume. However, since stars form only in the nested-grid region in our simulations, we need to estimate the normalized star formation rate based on this region.
This allows us to compare trends for different star formation models.}
First, the model of \citet{Kravtsov2003} (K03) that assumes that star formation is proportional to the local gas density. This assumption corresponds to a global Schmidt law.\footnote{To reduce fluctuations, we modified the code to support stochastic star formation also for the K03 model.}
The coefficient of proportionality is given by the inverse of a time scale, which we set to $1\,\mathrm{Gyr}$.
With a decline by about one order of magnitude toward low redshift, see dot-dashed orange line in Fig.~\ref{fig:sfr}, the K03 model is qualitatively in agreement with the observed redshift dependence of the star formation rate \citep{Behroozi2013}.
Nevertheless, the model breaks down close to $z=0$, as reflected by the sudden burst of star formation.
At intermediate redshifts ($z\sim1$), the KO3 model roughly agrees with an expected star formation efficiency $\varepsilon=0.01$.
The second model by \citet{Cen1992} (CO92) directly employs a stochastic star formation recipe with
$\varepsilon$ as free parameter controlling the star formation efficiency.
The two test runs with $\varepsilon=0.01$ and $\varepsilon=0.1$ (dotted blue and turquoise lines in Fig.~\ref{fig:sfr}) span the range of efficiencies that can be reasonably assumed.
However, both runs result in unrealistically high star formation at lower redshifts.
The third model we tested is the dynamical model proposed by \citet{Semenov16} (S+16).
In this case, the computation of the star formation efficiency is based on the local turbulent Mach number following from the SGS model.
While the model seems to perform well in simulations of isolated disk galaxies, our results suggest that the model initially overproduces stars, followed by a too rapid decline, see green dotted line in Fig.~\ref{fig:sfr}.
Apart from the underlying assumptions of the model, a potential problem is that one would need to scale the turbulent energy down to the scale of star forming clouds, which is much below the spatial resolution scale of our simulations.

Overall, the K03 model follows most closely the observed star formation rates. Thus, we decided to use the K03 model in our fiducial run. Finally, the threshold density above which gas can be turned into star particles needs to be adjusted to the numerical resolution. While we achieved good agreement for two and three AMR levels (see dashed and dot-dashed lines in Fig.~\ref{fig:sfr}), we increased the threshold by a factor of ten to a number density of $50\,\mathrm{cm}^{-3}$ in the case of the highest resolution. Otherwise, the star formation rate for $z\lesssim 2$ would have become too high, however, at the cost of a systematically lower star formation rate in the high-redshift regime.

The K03 model employs supernova feedback similar to the prescription of CO92. Basically, a certain fraction $f_{m\ast}$ of the stellar mass is assumed to be ejected, resulting in momentum feedback. Thermal feedback is controlled by the fraction $f_\mathrm{SN}$ of a star particle's rest energy that is deposited as thermal energy into the gas. The only difference compared to CO92 is that feedback is applied instantaneously, which is a reasonable approximation for resolutions in the kpc range, where the numerical time step $\Delta t\gtrsim 10^6\,\mathrm{yr}$ is comparable to or greater than the life time of massive stars. Momentum feedback potentially contributes to the production of turbulence, although gas motions that are induced locally at the grid scale will be strongly affected by numerical dissipation. Thermal feedback produces hot bubbles, particularly during episodes of intense star formation. The expansion of such bubbles can also give rise to turbulence. For this work, we adopt the commonly used default parameters $f_{m\ast}=0.25$ and $f_\mathrm{SN}=10^{-5}$. For further details, see \citet{Enzo2014} and references therein.
%__________________________________________________________________

\begin{figure*}
   \centering
   \includegraphics[width=\textwidth]{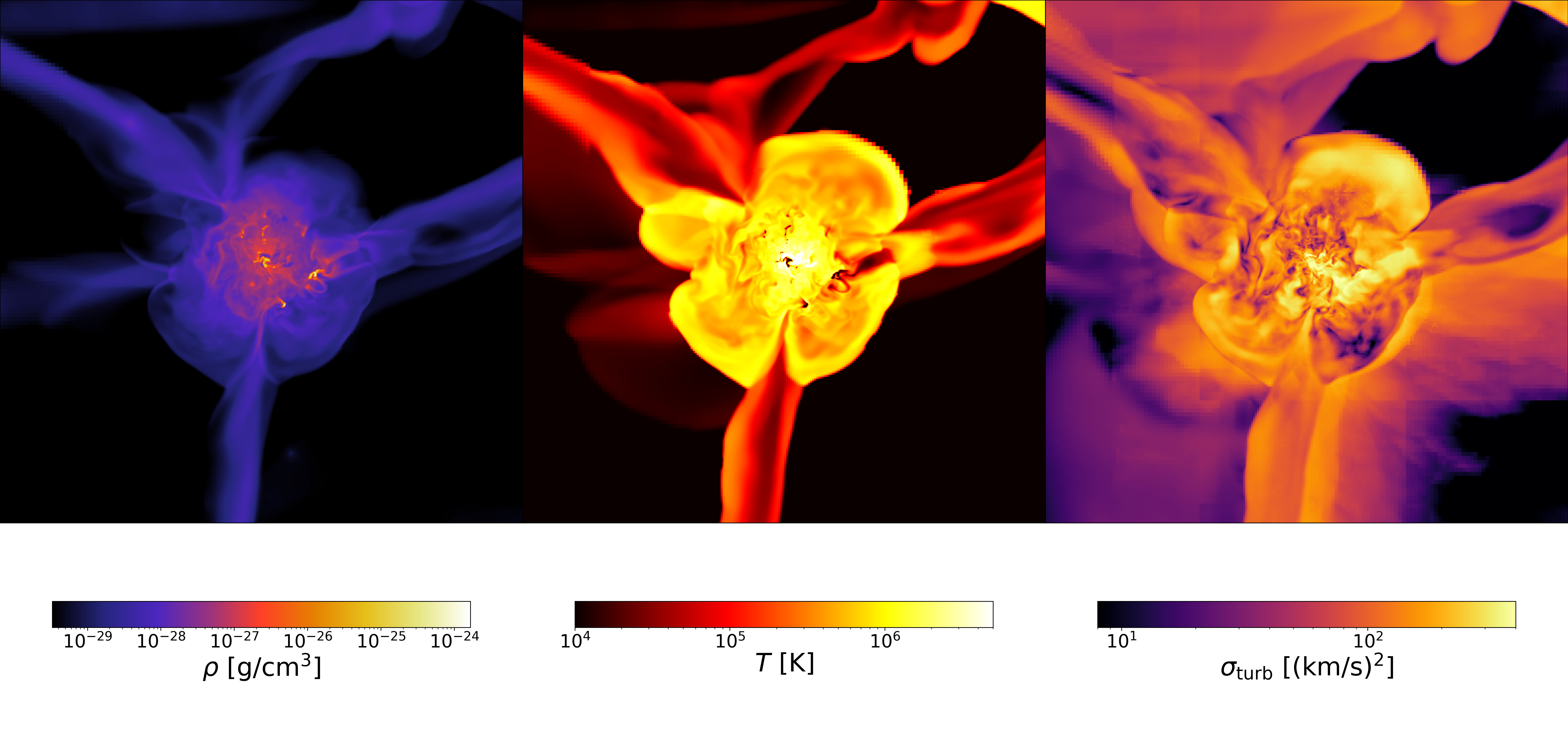}
   \caption{Slices of gas density, temperature, and turbulent velocity for the example compact group halo with a halo mass of $M_{\rm halo}=1.17\times 10^{12}M_\odot$ at redshift $z=1$. The panels are scaled to a size of $4R_{\rm vir}$, where the virial radius of the halo is $R_{\rm vir}=298 \,$kpc.}
   \label{fig:slices_z1}
\end{figure*}

\section{Results}
\label{sec:results}

In this section, we begin with a phenomenological discussion of individual halos accompanied by an analysis of halo profiles. This is the basis for determining constraints to distinguish the CGM/IGrM from other gas phases. Then we analyze mean CGM/IGrM values for all halos in the zoom-in region. We concentrate on the the cosmological epoch after the observed peak of star formation and AGN activity. To study evolutionary trends, statistics for $z=0$ and $1$ are compared.

\subsection{Individual halos}
\label{sec:profiles}

In the following, we analyze data from our high-resolution runs (4 nested grid levels, 4 AMR levels, global Schmidt law for number densities $n>50\,\mathrm{^{-3}}$). We applied the HOP finder of yt, to identify halos in the zoom-in region  \citep{Eisenstein98}. The HOP algorithm defines halos by searching for density peaks and grouping particles in distinct groups based on nested density contours (peak, saddle, and outer boundary). The resulting halos can be either isolated galaxies or groups of galaxies.\footnote{For a brief outline of the algorithm, see \url{https://yt-project.org/doc/analyzing/analysis\_modules/halo\_catalogs.html?highlight=hop\#hop}. We applied the halo finder with default parameters.}
At redshift $z=1$, 102 objects with halo masses between $10^{10}\,M_\odot$ and $1.32\times 10^{12}\,M_\odot$ and virial radii in the range from $28$ to $298\,\mathrm{kpc}$ were identified.\footnote{The halo mass is defined by the total mass of the particle group identified as halo by the HOP finder.} Halos of lower mass are excluded from our analysis. Apart from relatively poor mass resolution, turbulent structures smaller than about ten times the grid scale (i.e., $\lesssim 10\,\mathrm{kpc}$ at the maximum refinement level) are strongly damped by numerical dissipation \citep{Grete2017b}, making estimates of turbulent velocities in small, low-mass halos infeasible.  

\begin{figure*}
   \centering
   \includegraphics[width=\textwidth]{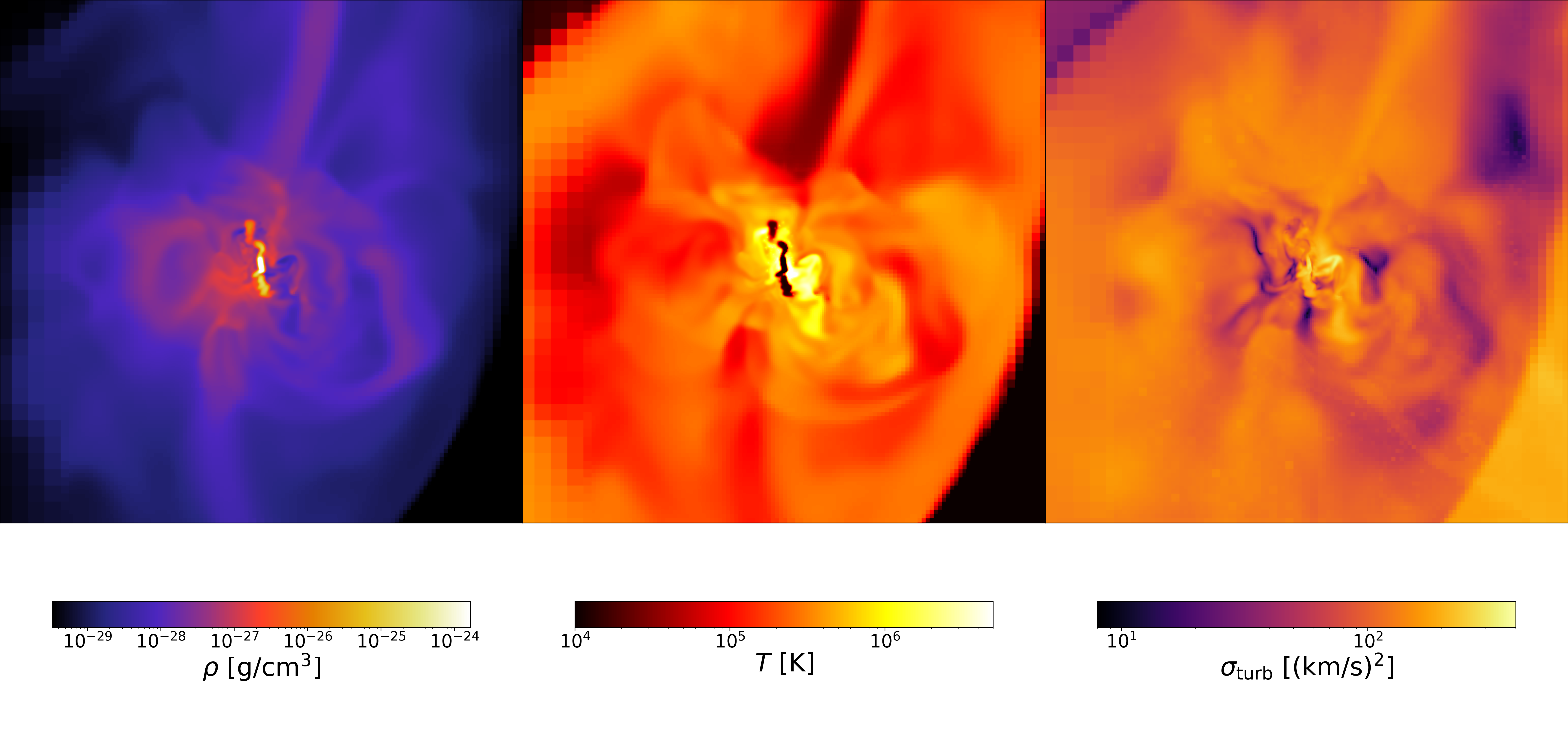}
   \caption{Same as in Fig.~\ref{fig:slices_z1} for the isolated galaxy example within a lower mass halo
   %($M_{\rm halo}=1.74\times 10^{11}M_\odot$, $R_{\rm vir}=467\,$kpc) at redshift $z=0$.
   ($M_{\rm halo}=9.90\times 10^{10}M_\odot$, $R_{\rm vir}=73\,$kpc).}
   % halos_3: mass = 1.220777e+12 Msun, radius = 434 kpc
   \label{fig:slices_z1_low_mass}
\end{figure*}

Examples are visualized in Figs~\ref{fig:slices_z1} and~\ref{fig:slices_z1_low_mass}, respectively, which show slices of the baryonic gas density $\rho$, the temperature $T$ of the gas, and the turbulent velocity dispersion $\sigma_{\rm turb}$ defined by equation~(\ref{eq:sigma_turb}). For a particular object, the slices are scaled to twice the diameter of the halo. The object shown in Fig~\ref{fig:slices_z1} is located outside of the major filament in the zoom-in region. Its mass of $1.17\times 10^{12}\,M_\odot$ is typical for a galactic halo. Two off-center density peaks in the lower right quarter of the region shown in Fig~\ref{fig:slices_z1} are correlated with low temperatures $T\lesssim 10^{4}\,\mathrm{K}$. These structures can be interpreted as satellites of the galaxy at the center of the halo. Within this halo there are sharp outward drops in density and temperature are associated with outer accretions shocks. As discussed in \citet{Schmidt2016}, the volume inside of the accretions shocks is filled with turbulent gas. Compared to the medium in the void, $\sigma_{\rm turb}$ is at least two orders of magnitude larger in the WHIM and ICM of clusters. For the halo shown in Fig.~\ref{fig:slices_z1}, the gas outside the accretion shocks is comparatively quiescent, but the change of $\sigma_{\rm turb}$ across the outer shocks is less pronounced (see right panel). The residuals of the order of $10\,\mathrm{km/s}$ in low-density regions are caused by slowly decaying temporal correlations in the filtered flow\footnote{See \citet{Schmidt2014}, section~3, for a detailed discussion.} Nevertheless, these residuals are typically a factor of ten smaller than the turbulent velocity dispersion inside the halo (see radial profiles below). 

\begin{figure*}
   \centering
   \includegraphics[width=\textwidth]{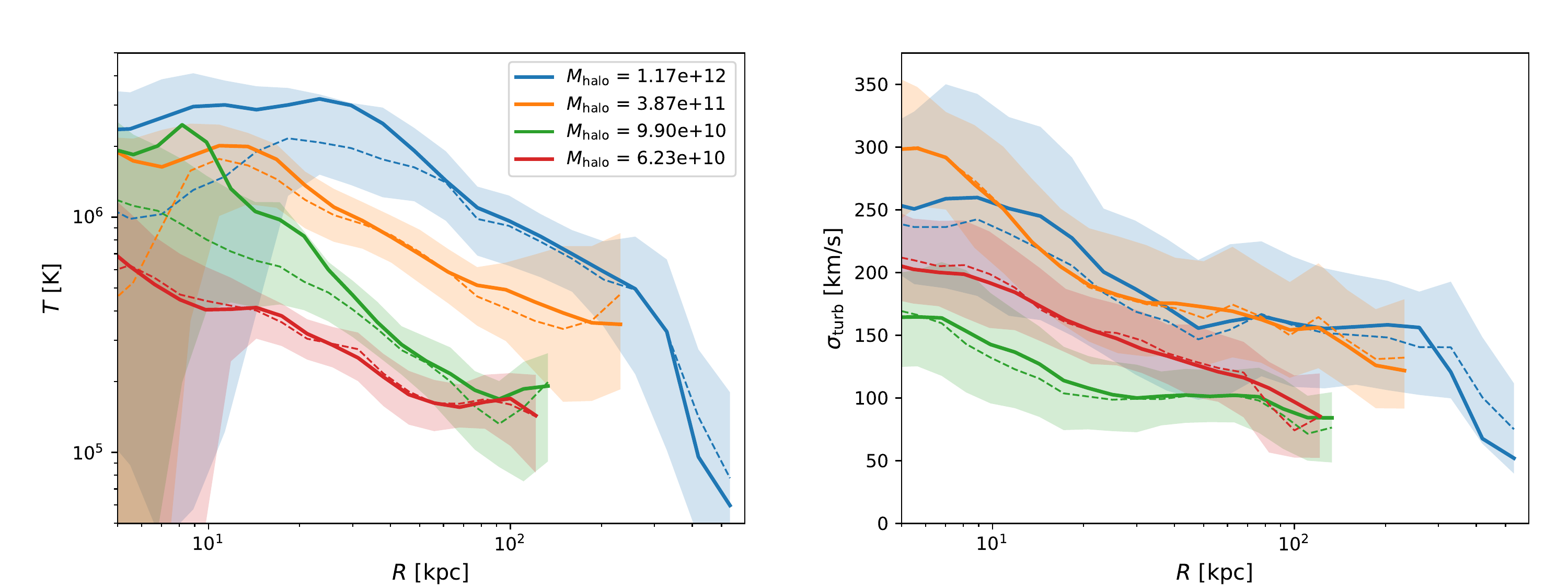}
   \caption{Radial profiles of temperature $T$ and turbulent velocity dispersion $\sigma_{\rm turb}$ for selected halos at redshift $z=1$. (Mean values per radial bin are shown as solid lines, medians as dashed lines, and interquartile ranges as shaded regions.)}
   \label{fig:profiles_z1}
\end{figure*}

The object shown in Fig.~\ref{fig:slices_z1_low_mass} has a halo mass that is by one order of magnitude smaller than in the previous example. At the center is a single galaxy with outflows that can be discerned as hot gas (middle slice) at low densities (left slice). In this case, a turbulent velocity dispersion of a few $100\,\mathrm{km/s}$ is found inside the galaxy and in some of the ejected gas. In most of the CGM, $\sigma_{\rm turb}$ is around $100\,\mathrm{km/s}$. This is confirmed by the radial profiles shown in Fig.~\ref{fig:profiles_z1}, where this halo (green line) is compared to the halo discussed above (blue line) and two additional halos with masses of $3.87\times 10^{11}\,M_\odot$ (orange line) and $6.23\times 10^{10}\,M_\odot$ (red line). While the gas of the most massive halo has profiles that resemble those of small galaxy clusters, i.e.\ they are relatively flat near the core and fall off steeply at the outskirts \citep{Schmidt2016}, the lower mass objects have pronounced peaks at the center and falls off more gradually with radius. As suggested by the profiles of the two additional halos, the transition is rather gradual and the shapes of the profiles vary substantially. It is important to keep in mind that scales $\lesssim 10$\,kpc are affected by numerical dissipation and that radial binning presumes spherical symmetry. Especially for galactic halos, the disk-like structures can strongly distort the profiles. For example, the central dip of the temperature profile for the halo with mass $9.9\times 10^{10}\,M_\odot$ is not as distinct as one would suppose, given the temperature slice shown in Fig.~\ref{fig:slices_z1_low_mass}. This is a consequence of averaging over low-temperature gas in the ISM and hot gas in the CGM in spherical shells. Separately, we assess the robustness of the profiles in Fig.~\ref{fig:profiles_z1} by plotting the mean and median values (solid vs.\ dashed lines) as well as the interquartile ranges (IQRs as shaded regions). In addition to the raw IQR, deviations between the mean and median indicate a large spread in a quantity. Those are particularly useful for quantities that vary over several orders of magnitude as the mean values tend to be dominated by the largest values in the sample. Beyond the central regions ($R\gtrsim10$\,kpc) the profiles exhibit a limited spread hinting at some first trends. Roughly speaking, more massive halos typically also have higher temperatures whereas there is no clear trend in the turbulent velocity dispersion for halo with different masses.

\begin{figure*}
   \centering
   \includegraphics[width=\textwidth]{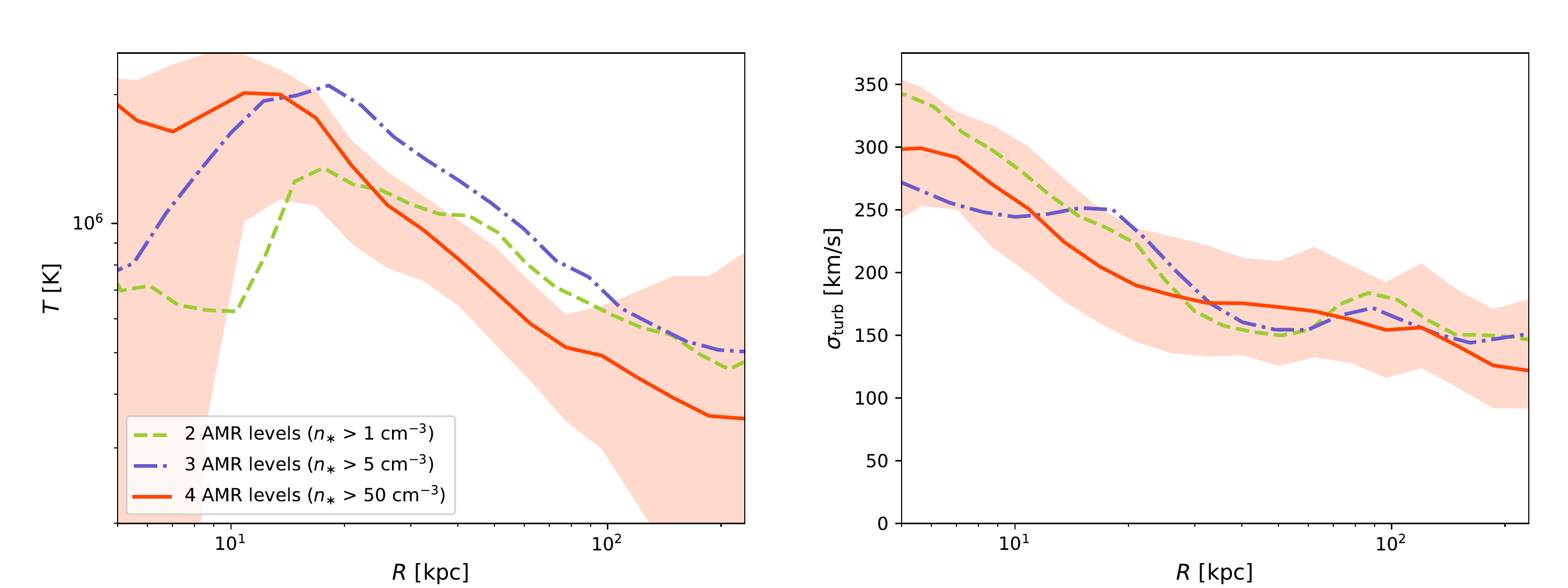}
   \caption{Resolution dependence of radial profiles (mean values) for the halo of mass mass $M_{\rm halo}=3.87\times 10^{11}M_\odot$ (see also Figs~\ref{fig:slices_z1_low_mass} and~\ref{fig:profiles_z1}). The shaded regions indicate the interquartile range for the highest resolution.}
   \label{fig:profiles_z1_res}
\end{figure*}

As discussed in Sect.~\ref{sec:numerics}, star formation and radiative cooling are sensitive to numerical resolution. The tuning of star formation and feedback model parameters to numerical resolution is a common problem in simulations of galaxy evolution. As an example, the impact of the number of refinement levels on radial profiles is shown in Fig.~\ref{fig:profiles_z1_res} for a halo of intermediate mass (i.~e., $M_\mathrm{halo}\sim 10^{11}\,M_\odot$). While the turbulent velocity dispersion profiles are comparable for different spatial resolutions, there are clearly deviations in the temperature profiles. At lower resolution, star formation activity is reduced and feedback becomes less efficient, resulting in a drop of the gas temperature in the core. At least the maximum temperatures agree for spatial resolutions of $2\,\mathrm{kpc}$ (3 AMR levels) and $1\,\mathrm{kpc}$ (4 AMR levels). With a radius smaller than a tenth of the virial radius, the core region fills only a small fraction of the total volume of halo. As a result, resolution effects can be expected to affect mass averages more strongly than volume averages.

\begin{figure*}
   \centering
   \includegraphics[width=\textwidth]{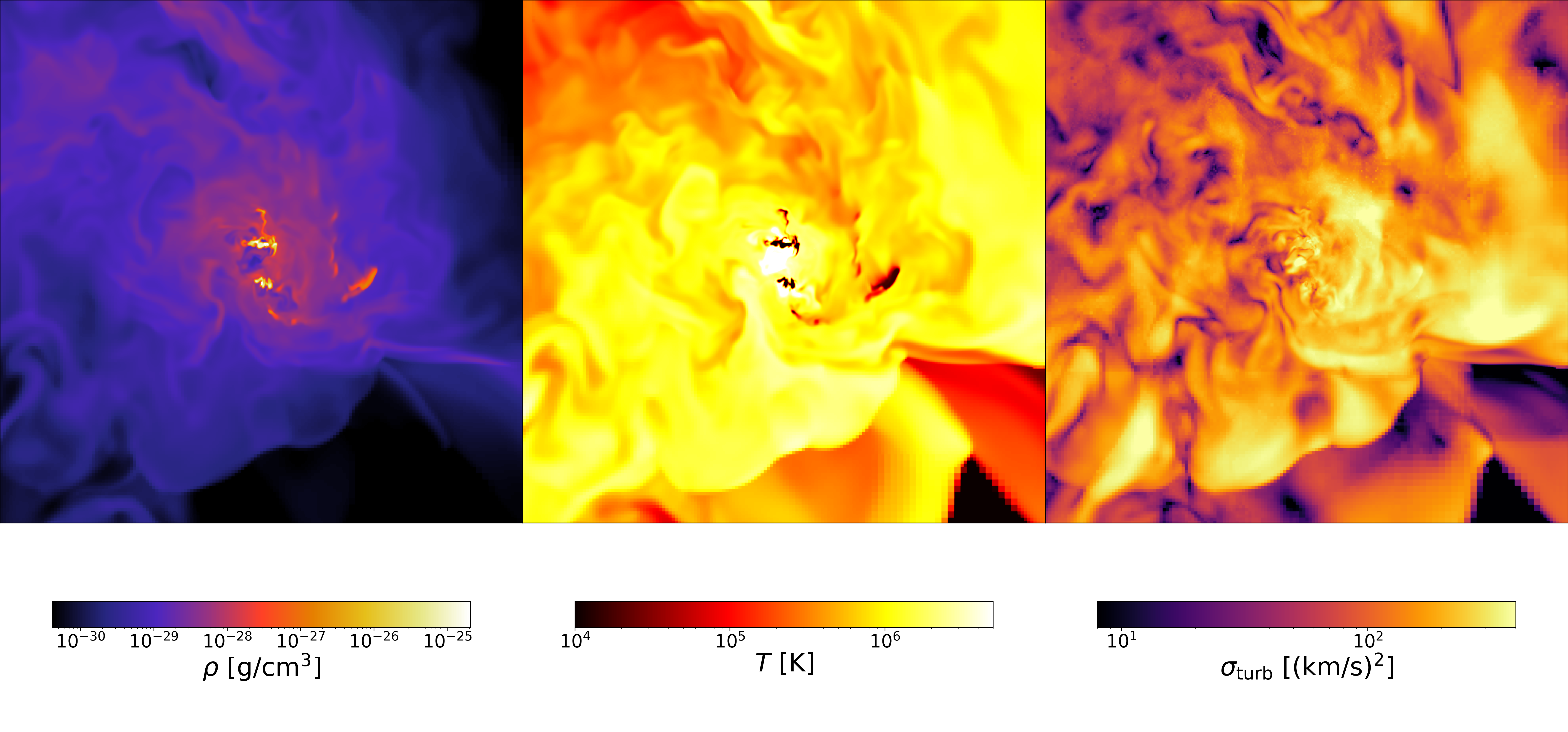}
   \caption{Same as in Fig.~\ref{fig:slices_z1} for a halo of 
   mass $M_{\rm halo}=1.58\times 10^{12}M_\odot$ and $R_{\rm vir}=373\,$kpc 
   at redshift $z=0$.}
   \label{fig:slices_z0}
\end{figure*}

\begin{figure*}
   \centering
   \includegraphics[width=\textwidth]{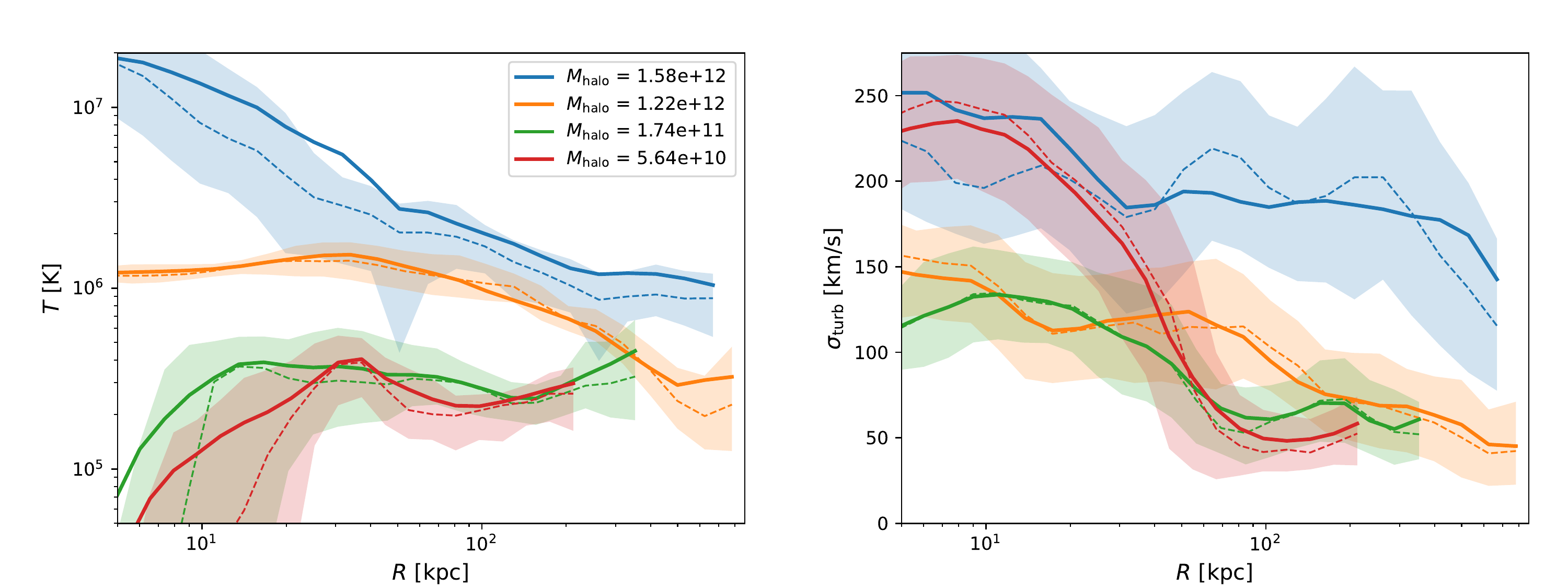}
   \caption{Radial profiles of temperature $T$ and turbulent velocity dispersion $\sigma_{\rm turb}$  selected halos at redshift $z=0$. (Mean values per radial bin are shown as solid lines, medians as dashed lines, and interquartile ranges as shaded regions.)}
   \label{fig:profiles_z0}
\end{figure*}

At redshift zero, the largest halo in the nested-grid region has a mass of $3.61\times 10^{12}\,M_\odot$ and a virial radius $R_{\rm vir}=467\,\mathrm{kpc}$. In total, we found 62 objects above $10^{10}\,M_\odot$ with virial radii greater than $50\,\mathrm{kpc}$. From the four halos with masses greater than $10^{12}\,M_\odot$, two are massive galaxies, while the others are composed of multiple objects. An example is shown in Fig.~\ref{fig:slices_z0}. There are two galaxies close to the center that are surrounded by relatively dense and hot gas. Quite likely, these galaxies are interacting. There are also smaller blobs of dense, cool gas that could be either residuals from an interaction or smaller satellite galaxies. Thus, it can be interpreted as a compact group at the lower end of the mass range of groups. In contrast to the halo of comparable mass shown in Fig.~\ref{fig:slices_z1}, the transition to the intergalactic medium of the filament is rather gradual. This becomes clear when looking at the temperature profile of this halo, see blue lines in Fig.~\ref{fig:profiles_z0}. The temperature changes only little with radius, except for the temperature increase in the core region. The profile of the turbulent velocity dispersion is also rather flat. The halo of mass $1.22\times 10^{12}\,M_\odot$ (orange lines) is an isolated, massive galaxy. There is no drop in temperature in the outskirts either and the temperature profile is flat throughout the halo. This object might be a fossil group, which is the end state of former group members merging into a single, dominant galaxy. The other profiles show two single galaxies (green and red lines) with halo masses $\sim 10^{11}\,M_\odot$. They are dominated by cooler gas close to the center. Interestingly, they exhibit strong turbulence at radii $\sim 10\,\mathrm{kpc}$, where $\sigma_\mathrm{turb}$ is in the range from $100$ to more than $200\,\mathrm{km/s}$. Further outside in the CGM, $\sigma_\mathrm{turb}$ decreases to a significantly lower background level. This suggests that feedback from the galaxies enhances turbulence.

\subsection{Halo statistics}
\label{sec:results_halos}

The key element of our analysis is the computation of mean energies for all halos in the zoom-in region with mass $M_{\rm halo}~>~10^{10}M_\odot$. The standard method is to average over all cells within the virial sphere, i.e.\ from the center of the halo to the radius $R_{\rm vir}$. However, our phenomenological discussion in Sect.~\ref{sec:profiles} shows that warm, tenuous gas can be found inside the virial radius, while hotter gas at higher density can extend beyond the virial radius. For this reason, we need additional constraints to distinguish the CGM or IGrM from other gas phases in filaments.

\subsubsection{Identifying the CGM/IGrM}
\label{sec:constraints}
To remove the warm, tenuous gas we exclude gas below a minimum overdensity $\delta_\mathrm{min}$ relative to the mean density of baryonic and dark matter and below a minimum temperature $T_\mathrm{min}$:
\begin{equation}
    \label{eq:constr}
    \delta > \delta_\mathrm{min}, \quad \mathrm{and} \quad T > T_\mathrm{min}.
\end{equation}
In addition, we only include gas with a neutral hydrogen number density below $1\,\mathrm{cm^{-3}}$.
These constraints are applied to a spherical volume of radius $R_\mathrm{max}=2R_\mathrm{vir}$. On the one hand, this includes gas in the outskirts outside of the virial sphere. On the other hand, dense gas at low temperatures inside galaxies, i.e., star-forming gas, is excluded. A maximal radius of $2R_\mathrm{vir}$ corresponds to the regions shown Figs~\ref{fig:slices_z1}, \ref{fig:slices_z1_low_mass}, and~\ref{fig:slices_z0}. It is sufficiently large to contain structures like the galactic halo in Fig.~\ref{fig:slices_z1_low_mass} while a significant overlap with neighboring halos is avoided.

\begin{figure*}
   \centering
   \includegraphics[width=\textwidth]{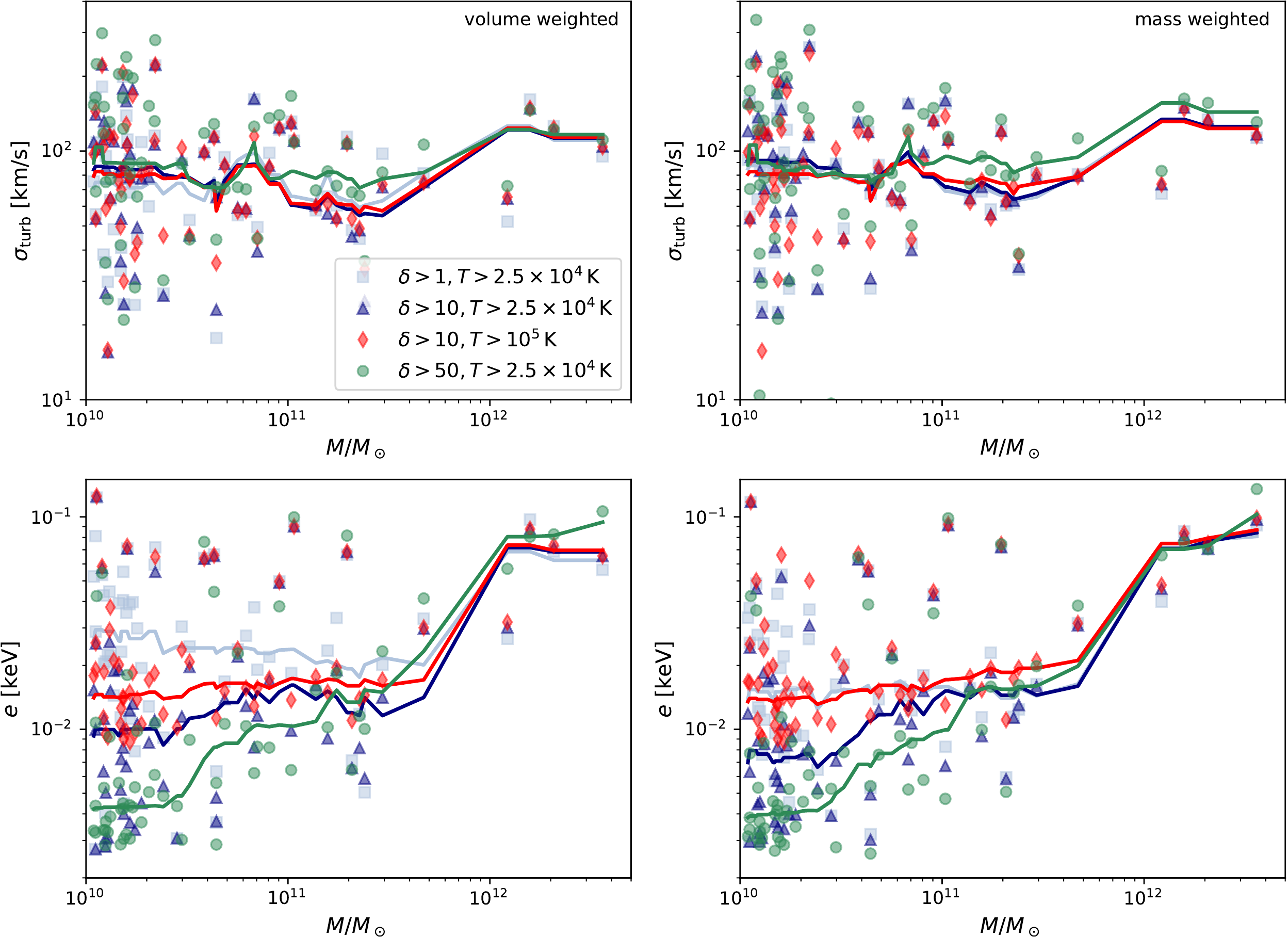}
   \caption{Volume-weighted (left) and mass-weighted (right) mean values of the turbulent velocity dispersion and thermal energy of halos for different constraints at $z=0$. Solid lines show medians of objects in the mass range $[M/2,2M]$ for each halo of mass $M$.}
   \label{fig:mean_weighing}
\end{figure*}

To quantitatively evaluate different thresholds, we first computed the volume- and mass-weighted averages of the turbulent velocity dispersion and thermal energy of all halos at redshift $z=0$ for varying overdensity $\delta_\mathrm{min}$ of  1, 10, and 50 at a fixed temperature limit $T_\mathrm{min}=2.5\times 10^4\,\mathrm{K}$, see Fig.~\ref{fig:mean_weighing}. While the chosen thresholds are arbitrary within the range of plausible densities and temperatures of the CGM/IGrM, they avoid assumptions such as spherical symmetry or correlations between the properties of a dark-matter halo and its gas contents (see also the discussion in \citealt{Schmidt2016} in the context of clusters). The turbulent velocity dispersion $\sigma_\mathrm{turb}$ exhibits no clear trend with halo mass, except that the scatter increases toward low-mass halos. The results are overall not particularly sensitive to the chosen threshold density. As indicated by the sliding median for a mass window $[M/2,2M]$ (solid lines in Fig.~\ref{fig:mean_weighing}), the typical turbulent velocity dispersion is of the order of $100\,\mathrm{km/s}$. The mean thermal energies suggest a drop from halo masses above $M_{\rm halo}\sim 10^{12}M_\odot$ to lower masses. Since the four most massive halos are both galaxies and small groups, it is unclear whether this drop has any significance or is merely a statistical fluctuation. Similar to the turbulent velocity dispersion, we find larger scatter for halos of lower mass. Moreover, some trends with the thresholds $\delta_\mathrm{min}$ and $T_\mathrm{min}$ can be discerned: Toward the low-mass end, thermal energies tend to be higher for lower $\delta_\mathrm{min}$. This can be understood as a consequence of excluding high-temperature gas at relatively low densities, for example, the shocked gas that can be seen in Fig.~\ref{fig:slices_z1_low_mass}. As expected, this trend is reduced for the lowest density threshold if mass weighing is applied. We also calculated the mean turbulent velocity dispersion and thermal energy for a higher minimum temperature of $T_\mathrm{min}=5\times 10^4\,\mathrm{K}$ at intermediate overdensity $\delta_\mathrm{min}=10$ to assess the effect of $T_\mathrm{min}$. While the overall impact of the temperature threshold is rather minor, one can see a clustering of the mean thermal energies just above the chosen temperature threshold. This means that objects with a significant fraction of gas in the temperature range below the threshold are shifted upwards, which should be avoided. On the other hand, choosing $T_\mathrm{min}$ significantly lower than $2.5\times 10^4\,\mathrm{K}$, would partially mix up the surroundings of galaxies with warm component of the ISM.

\begin{figure*}
   \centering
   \begin{subfigure}{0.5\textwidth}
      \includegraphics[width=\linewidth]{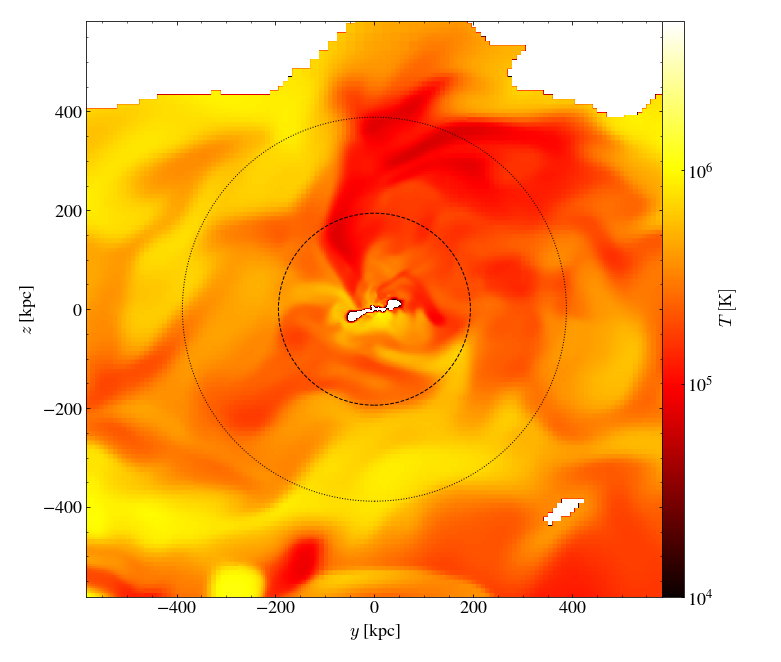}
      \subcaption{$\delta>1$, $T>2.5\times 10^4\unit{K}$}
   \end{subfigure}\hfill
   \begin{subfigure}{0.5\textwidth}
      \includegraphics[width=\linewidth]{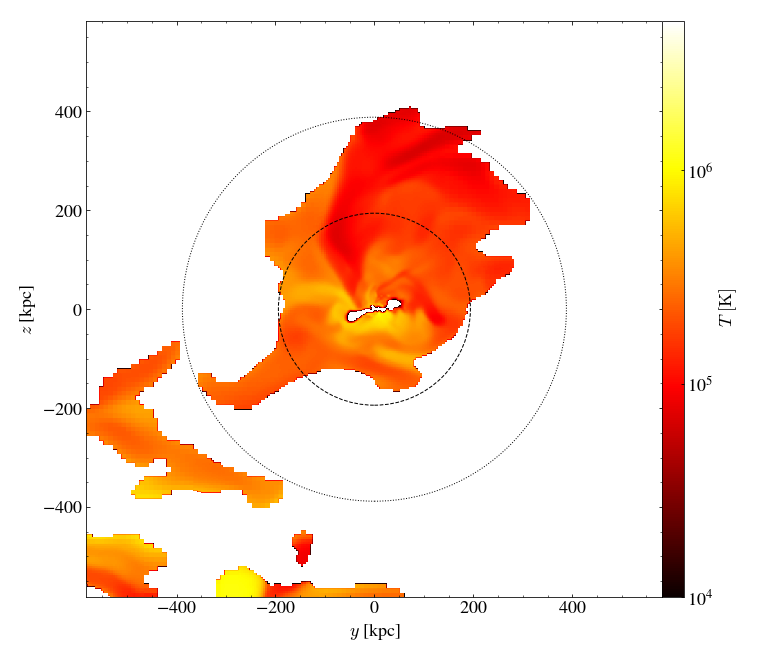}
      \subcaption{$\delta>10$, $T>2.5\times 10^4\unit{K}$}
   \end{subfigure}\hfill
   \begin{subfigure}{0.5\textwidth}
     \includegraphics[width=\linewidth]{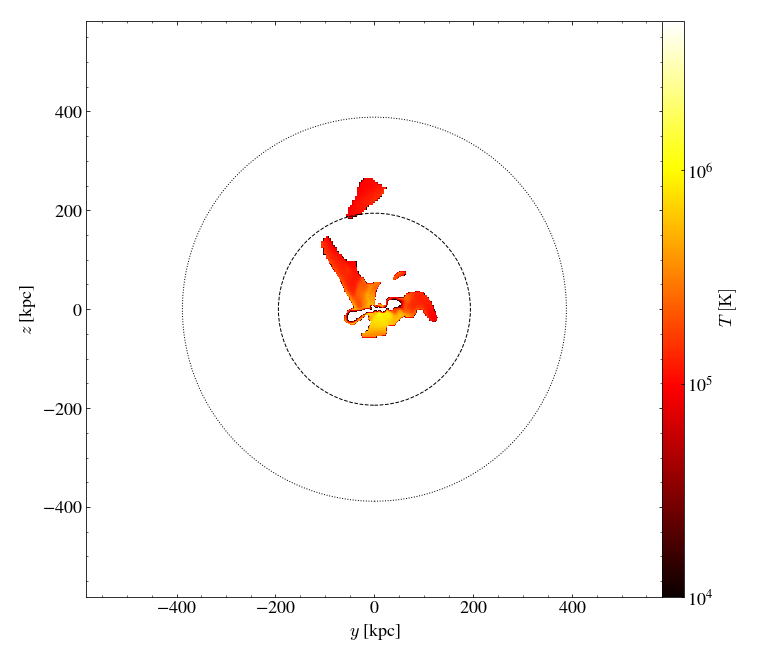}
     \subcaption{$\delta>50$, $T>2.5\times 10^4\unit{K}$}
   \end{subfigure}\hfill
  \begin{subfigure}{0.5\textwidth}
     \includegraphics[width=\linewidth]{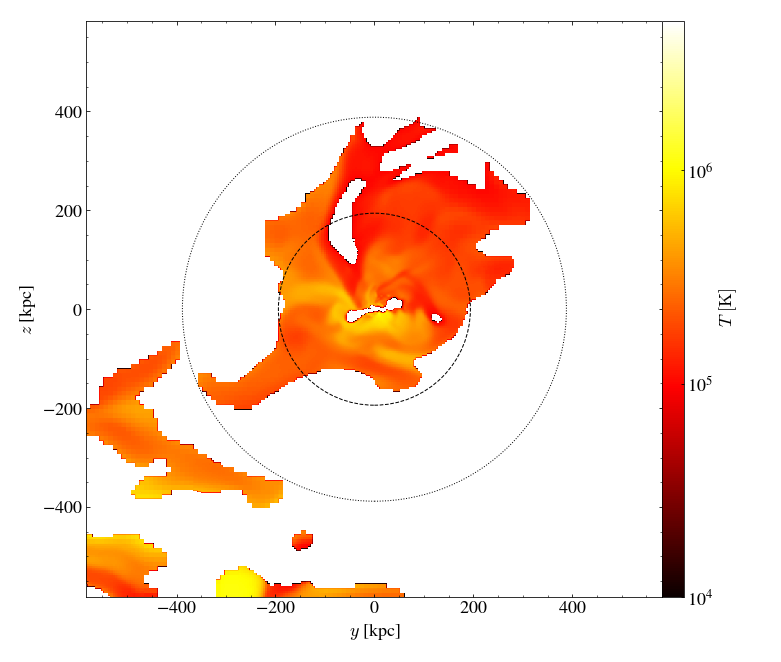}
     \subcaption{$\delta>10$, $T>10^5\unit{K}$}
  \end{subfigure}\hfill
   \caption{Temperature slices constrained by overdensity $\delta > \delta_\mathrm{min}$ and temperature $T > T_\mathrm{min}$ for a halo of mass $M_{\rm halo} = 1.74\times 10^{11} M_\odot$ at redshift $z=0$. The inner circle shows the virial radius $R_{\rm vir}=194\,$kpc of the halo. For the computation of halo averages based on these constraints, a maximal radius of $2R_{\rm vir}$ (outer circle) is applied.}
   \label{fig:slices_def}
\end{figure*}

To gain a qualitative understanding of the thresholds, Fig.~\ref{fig:slices_def} shows the gas temperature inside the regions constrained by the $\delta_\mathrm{min}$ and $T_\mathrm{min}$ for the halo with $M_{\rm halo} = 1.74\times 10^{11} M_\odot$ at $z=0$ (see also Fig.~\ref{fig:profiles_z0}). In the case $\delta_\mathrm{min}=1$ and $T_\mathrm{min}=2.5\times 10^4\,\mathrm{K}$ (top left panel in Fig.~\ref{fig:slices_def}), only the ISM of the galaxy in the center is excluded and the region for which the average is computed is simply cut off at $r=2R_{\rm vir}$. The opposite extreme is $\delta_\mathrm{min}=50$ $T_\mathrm{min}=2.5\times 10^4\,\mathrm{K}$ (bottom left), where only the most dense part of the CGM is included. The intermediate threshold ($\delta_\mathrm{min}=10$, right panels) effectively constrains the CGM to gas inside $2R_{\rm vir}$. In agreement with the preceding analysis, the resulting mean values of the thermal energy become higher if gas at lower densities is included (see Table \ref{tab:weighing_means}), while the differences between $T_\mathrm{min}=2.5\times 10^4\,\mathrm{K}$ (top right) and $10^5\,\mathrm{K}$ (bottom right) are rather small for this halo. The largest turbulent velocity dispersion is found for $\delta_\mathrm{min}=50$. In this case, turbulent gas in the close vicinity of the galaxy fills a relatively large volume fraction. As indicated by the radial profile plotted in Fig.~\ref{fig:profiles_z0}, the mean turbulent velocity dispersion tends to decrease if more gas from regions at radii $\gtrsim 100\,$kpc contributes to the average.

\begin{table}[]
    \caption{Mean thermal energy and turbulent velocity dispersion of the halo shown in Fig.~\ref{fig:slices_def} for different density and temperature thresholds.}
    \label{tab:weighing_means}
    \centering
    \begin{tabular}{rrcc}
    \hline\hline    
    $\delta_\mathrm{min}$ & $T_\mathrm{min}$ & $\langle\sigma_\mathrm{turb}\rangle\,$[km/s] &
    $\langle e\rangle\,$[keV] \\
    \hline
$1$ & $2.5\times 10^4\,$K & 60.5 & 0.0334 \\
$10$ & $2.5\times 10^4\,$K & 53.9 & 0.0190 \\
$10$ & $10^5\,$K & 53.3 & 0.0195 \\
$50$ & $2.5\times 10^4\,$K & 72.6 & 0.0184 \\
    \hline
    \end{tabular}
\end{table}

\begin{figure*}
   \centering
   \includegraphics[width=\textwidth]{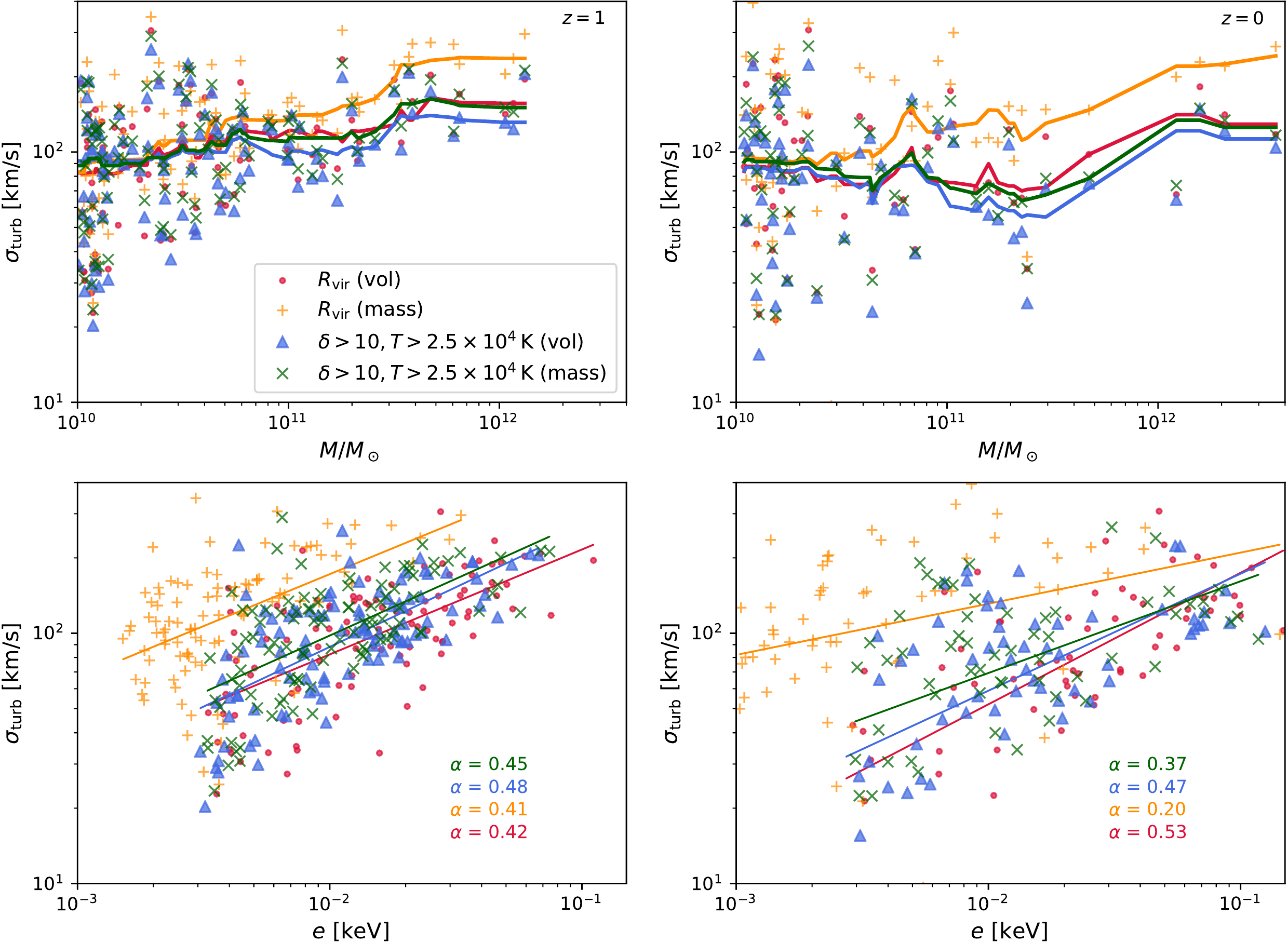}
   \caption{Mean values of the turbulent velocity dispersion vs halo mass (top) and mean thermal energy (bottom) for halos at redshift $z=1$ (left) and $z=0$ (right). As indicated in the legend, both volume- and mass-weighted averages are computed for cells within the virial radius or cells with density and temperature above the specified thresholds. While the thick solid lines in the top plots ($\sigma_\mathrm{turb}$ vs $M$) are sliding medians as Fig.~\ref{fig:mean_weighing}, the lower plots ($\sigma_\mathrm{turb}$ vs $e$) show power-law fits with slope $\alpha$.}
   \label{fig:mean_evol}
\end{figure*}

\begin{table}[]
    \caption{Median values of the turbulent velocity dispersion for $M_{\rm halo} \ge 10^{10}M_\odot$.}
    \label{tab:sigma_median}
    \centering
    \begin{tabular}{clr}
    \hline\hline    
    $z$ & region (weighing) & median [km/s] \\
    \hline
    $1$ & $R_\mathrm{vir}$ (vol) &  105.3\\ 
    $1$ & $R_\mathrm{vir}$ (mass) & 121.3\\
    $1$ & $\delta > 10, T > 2.5\times 10^4\,$K (vol) & 100.7\\
    $1$ & $\delta > 10, T > 2.5\times 10^4\,$K (mass) & 110.2\\
    $0$ & $R_\mathrm{vir}$ (vol) & 83.3\\
    $0$ & $R_\mathrm{vir}$ (mass) & 120.2\\
    $0$ & $\delta > 10, T > 2.5\times 10^4\,$K (vol) & 80.5\\
    $0$ & $\delta > 10, T > 2.5\times 10^4\,$K (mass) & 84.2\\
    \hline
    \end{tabular}
\end{table}

\subsubsection{Scaling laws}
\label{sec:scaling}

Based on the quantitative and qualitative analysis, we we choose $\delta_\mathrm{min}=10$ and $T_\mathrm{min}=2.5\times 10^4\,\mathrm{K}$ for the following analysis.
Again, to evaluate the robustness of this choice we also compare the results to the ones obtained from directly averaging over the virial spheres. The resulting mean values of $\sigma_\mathrm{turb}$ at redshifts $z=1$ and $0$ are plotted in Fig.~\ref{fig:mean_evol}. Both definitions of averages result in similar distributions of $\sigma_\mathrm{turb}$ vs.\ halo mass, which is confirmed by the sliding medians. For the complete sample, however, we find that mass-weighted averages over virial spheres tend to be larger compared to other averages (see medians of $\sigma_\mathrm{turb}$ listed in Table~\ref{tab:sigma_median}). Since the radial profiles (see Figs~\ref{fig:profiles_z1} and~\ref{fig:profiles_z0}) show that $\sigma_\mathrm{turb}$ is larger in the cores than in the outskirts, mass-weighing emphasizes the peak values of $\sigma_\mathrm{turb}$. As shown in the preceding section, low-mass halos exhibit a stronger scatter. Moreover, at lower redshift this scatter becomes even more pronounced, suggesting that evolutionary effects beyond the self-similar gravitational collapse enhance the diversity of low-mass halos. The overall median values confirm a trend of decreasing $\sigma_\mathrm{turb}$ from $z=1$ to $0$, suggesting that energy injection is reduced toward lower redshift. This is in agreement with the decline of stellar feedback, which contributes to the production of turbulence in the surroundings of galaxies.

The strong scatter of the energy-mass relations is confirmed by the computation of correlation coefficients. Since the data cannot be assumed to follow normal distributions,\footnote{This is easily confirmed by plotting histograms of the data.} we use Spearman's nonparametric measure of correlation. The results are listed in Table~\ref{tab:correl}. While moderate correlations are found for $z=1$, the halo masses become weakly correlated or uncorrelated with thermal energy and turbulent velocity dispersion at $z=0$.

\begin{table*}[]
    \caption{Spearman’s correlations (correlation coefficient $r$ and $p$-value) between halo mass $M$, mean thermal energy $e$, and turbulent velocity dispersion $\sigma_\mathrm{turb}$ for the data shown in Fig.~\ref{fig:mean_evol}.}
    \label{tab:correl}
    \centering
    \begin{tabular}{clcccccc}
    \hline\hline
    & & \multicolumn{2}{c}{$e$--$M$} & \multicolumn{2}{c}{$\sigma_\mathrm{turb}$--$M$} &
    \multicolumn{2}{c}{$\sigma_\mathrm{turb}$--$e$} \\
    \hline
    $z$ & region (weighing) & $r$ & $p$ & $r$ & $p$ & $r$ & $p$\\
    \hline
    $1$ & $R_\mathrm{vir}$ (vol) & 0.482 & $2.9\times 10^{-7}$ & 0.450 & $2.1\times 10^{-6}$ & 0.638 & $5.3\times 10^{-13}$\\ 
    $1$ & $R_\mathrm{vir}$ (mass) & 0.390 & $5.0\times 10^{-5}$ & 0.591 & $6.0\times 10^{-11}$ & 0.538 & $5.7\times 10^{-9}$\\
    $1$ & $\delta > 10, T > 2.5\times 10^4\,$K (vol) & 0.424 & $9.0\times 10^{-6}$ & 0.329 & $7.3\times 10^{-4}$ & 0.654 & $8.8\times 10^{-14}$\\
    $1$ & $\delta > 10, T > 2.5\times 10^4\,$K (mass) & 0.482 & $3.0\times 10^{-7}$ & 0.395 & $3.9\times 10^{-5}$ & 0.641 & $3.9\times 10^{-13}$ \\
    $0$ & $R_\mathrm{vir}$ (vol) & 0.151 & 0.24 & 0.074 & 0.57 & 0.702 & $2.0\times 10^{-10}$\\
    $0$ & $R_\mathrm{vir}$ (mass) & 0.101 & 0.44 & 0.262 & 0.040 & 0.418 & $7.3\times 10^{-4}$\\
    $0$ & $\delta > 10, T > 2.5\times 10^4\,$K (vol) & 0.305 & 0.016 & -0.028 & 0.83 & 0.540 & $5.9\times 10^{-6}$\\
    $0$ & $\delta > 10, T > 2.5\times 10^4\,$K (mass) & 0.412 & $8.7\times 10^{-4}$ & 0.006 & 0.96 & 0.442 & $3.2\times 10^{-4}$ \\
    \hline
    \end{tabular}
\end{table*}

A different picture emerges when plotting $\sigma_\mathrm{turb}$ vs.\ the thermal energy $e$. Similar to the analysis of clusters in \citet{Schmidt2016}, we find a correlation between the turbulent and thermal energies (with $\sigma_\mathrm{turb}$ as proxy of the turbulent kinetic energy. The correlation coefficients listed in Table~\ref{tab:correl} are in the range between 0.5 and 0.7 for volume-weighted averages and somewhat lower of mass-weighted averages. The rather high correlations are confirmed by $p$-values: For $p<0.05$, it can be excluded at $95\,\%$ confidence level that data sets appear correlated by chance. However, compared to clusters (halo mass above $10^{13}\,M_\odot$), the scatter is stronger. Power-law fits are shown as straight lines in Fig.~\ref{fig:mean_evol} and their slopes $\alpha$ are listed in the lower legends. Although there are small deviations between the slopes following from constraints~\eqref{eq:constr} and volume-averaged virial spheres, the data basically agree within the scatter. Compared to mass-weighted averages over virial spheres, however, the discrepancy is large. By inspecting individual halos, it clearly follows that the mean thermal energy is shifted to systematically lower values if they are computed for all gas inside the virial radius. The explanation is quite simple: Mass-weighing emphasizes dense gas, including the cold gas in the interstellar medium. This results in a significant bias that is avoided by excluding gas at densities and temperatures that are characteristic for the ISM. These contributions are reduced if volume-weighing is applied. Apart from that, volume-averaging moves the focus away from the close vicinity of galaxies to the outskirts of the CGM/IGrM. For this reason, we calculated the slopes for the volume-averaged mean values based on the density and temperature thresholds at different redshift, $z=\{1,0.5,0.25,0\}$, see Table~\ref{tab:slope}. The resulting $\alpha\sim 0.47$ varies only little with redshift and is consistent with the scaling $\sigma_\mathrm{turb}\propto e^{\,0.5}$, i.e.\ a roughly constant turbulent Mach number at given redshift (see also \citealt{Schmidt2016}). However, the power-law coefficient $\sigma_0$, i.e. the turbulent velocity dispersion at an energy of $1\,\mathrm{keV}$, decreases with redshift.

\renewcommand{\arraystretch}{1.5}

\begin{table}[]
    \caption{Fit parameters for the power-law model
    $\langle\sigma_\mathrm{turb}\,[\mathrm{km/s}]\rangle = \sigma_0 \langle e\,[\mathrm{keV}]\rangle^\alpha$, where volume-weighted mean values are computed for overdensities $\delta > 10$ and $T > 2.5\times 10^4\,$K.}
    \label{tab:slope}
    \centering
    \begin{tabular}{lcc}
    \hline\hline    
    $z$ & $\alpha$ & $\sigma_0\,[\mathrm{km/s}]$ \\
    \hline
    $1$ & $0.477\pm 0.052$ & $793_{-166}^{+210}$ \\
    %a: 0.477 0.052 
	%b: 793 -166 210
    $0.5$ & $0.479\pm 0.063$ & $638_{-151}^{+198}$ \\
    %a: 0.479 0.063 
	%b: 638 -151 198
    $0.25$ & $0.473\pm 0.042$ & $522_{-81}^{+96}$ \\
    %a: 0.473 0.042 
	%b: 522 -81 96
    $0$ & $0.468\pm 0.079$ & $507_{-151}^{+214}$ \\
    %a: 0.468 0.079 
	%b: 507 -151 214
    \hline
    \end{tabular}
\end{table}

%__________________________________________________________________

\section{Conclusions}

\label{sec:conclusion}

We performed nested-grid simulations of a filament, applying AMR to increase the spatial resolution in halos of masses below $10^{13}\,M_\odot$. For these objects, which can be interpreted as galaxies and groups of galaxies, we analyzed the thermal and turbulent energy contents of the circumgalactic (CGM) and, in a few cases, the intragroup (IGrM) medium at the lower-mass end of groups. We applied a standard recipe for star formation (constant star formation time scale above a number density threshold) and supernova feedback.

Our study was motivated by the question whether the energy of the CGM/IGrM scales differently in the mass regime of groups or individual galaxies compared to halos of higher mass, i.e.\ in the range of clusters. Since we approached this question from a physical point of view, we chose hydrodynamical variables as metrics, namely, the thermal energy of the gas and the kinetic energy of turbulent gas flows. For the latter, we use the turbulent velocity dispersion as associated quantity. A meaningful definition of the turbulent velocity dispersion must be based on an integral quantity, encompassing velocity fluctuations on all scales. For this reason, we applied a Kalman filtering algorithm to estimate the numerically resolved component \citep{Schmidt2014} and the subgrid-scale model of \citet{Grete2016} for the unresolved component.

To infer scaling relations from the simulation data, we need to compute mean energies. The standard procedure is to apply a halo finder and to average over the virial sphere of each halo. However, the virial radius is only a crude way of specifying the boundaries of the gas belonging to a group of galaxies or the gas surrounding an isolated galaxy. For this reason, we investigated various criteria for defining the CGM based on density and temperature thresholds. We find a reasonable constraint when averaging over moderately overdense (a factor of 10 higher than the mean cosmological density) and warm-hot gas (above $2.5\times 10^4$\,K) within a maximum radius of two times the virial radius. The former excludes the cool-warm ISM inside galaxies whereas the latter avoids an overlap between halos. We find no clear trend of the turbulent velocity dispersion with halo mass, except for a larger scatter toward lower masses (top plots in Fig.~\ref{fig:mean_evol}). The scatter also tends to be more pronounced at lower redshift. The median for all halos in the nested-grid region is about $100\,\mathrm{km/s}$ at $z=1$ and $80\,\mathrm{km/s}$ at redshift zero (see Table~\ref{tab:sigma_median}). It turns out that ISM contributions cause significant deviations of the statistics inferred from mass-weighted averages using virial spheres in the mass range of galactic halos. In contrast, we find only minor differences between volume averages over virial spheres and regions constrained by density and temperature, which supports the robustness of our results. The power-law relation between turbulent velocity dispersion and thermal energy (bottom plots in Fig.~\ref{fig:mean_evol}) with a scaling exponent around $\alpha \sim 0.47$ at $z=0$ is similar to the relation found in simulations of galaxy clusters ($\alpha\sim 0.5$, see \citealt{Schmidt2016} and references therein). In other words, halos filled with hot gas also tend to be turbulent, regardless of the halo mass. This appears to apply all the way from the CGM to the IGrM to the ICM of massive clusters, although the scatter becomes larger with decreasing halo mass. Recently, \citet{Lochhaas2021} confirmed for an individual halo that the turbulent energy is a nearly constant fraction of the thermal energy over time.

Radial profiles of individual halos suggest that outflows from galaxies may produce high levels of turbulence in the close vicinity of star-forming galaxies, while the turbulent velocity dispersion decreases steeply in the outer regions of the CGM (see, for example, the sigma-shaped profile in Fig.~\ref{fig:profiles_z0}). Turbulence in the IGrM can be driven by mergers. In particular, fossil groups at low redshifts tend to be highly turbulent, as illustrated in Fig.~\ref{fig:slices_z0}. These effects result in substantially varying turbulence among different objects, obscuring the scaling relations found for halos of higher mass. Our results indicate that gravitational potential energy reservoir, for which the halo mass can be considered as proxy, becomes increasingly modulated by additional sources of energy in halos of lower mass. Tidal interactions between galaxies and supernova feedback during episodes of intense star formation both heat the surrounding gas and stir up turbulence, resulting in what is sometimes called second self-similarity \citep{Miniati2015,Schmidt2017}. Owing to their transient nature, interactions and feedback introduce stronger variations and, as a result, weaker correlation.

Although AGN feedback is not decisive for understanding the properties of the IGrM \citep{Liang2016}, it also has an impact on the galactic environment, particularly at earlier epochs ($z>1$). For this reason, incorporating AGNs into the subgrid physics will be an important component in improving the simulations discussed in this work.
Similarly, the CGM and IGrM are weakly magnetized \citep{Han2017}.
Like the ICM , the weakly collisional nature  of the plasma make it prone to fast growing kinetic
instabilities that (may) alter the magnetic field structure and saturation strength in a
turbulent environment\citep{Schekochihin2005}.
Thus, a treatment of magnetic fields in the simulations would further increase their fidelity.
In addition, it was recently shown that the standard adaptive refinement condition based on
dark matter and baryon overdensity, which we also used, (naturally) misses some details
in the structure of the low density outskirts \citep{Peeples2019,Hummels2019}.
In turn, this may alter the turbulent velocity dispersion at larger radii.
Given the increased amount resources required to resolve a larger fraction of the filament at
very high resolution, we leave this analysis to a future simulation campaign.

From an observational point of view, the situation is even more complicated because the X-ray luminosity is the main indicator of the dynamical state of the IGrM. Indeed, observed group luminosities point toward stronger scatter, but the question whether the scaling of X-ray luminosity vs temperature breaks at halo masses characteristic for groups is not settled \citep{Bharadwaj2014,Vajgel2014,Lovisari2015,Liang2016,Paul2017}. Although the emission of bremsstrahlung depends on $T^{1/2}$ and we found that $T$ scales with $\sigma_\mathrm{turb}$, the dependence on the squared number density of the gas might substantially weaken the relation between
X-ray luminosity and turbulent energy. In future work, it will be important to analyze the imprint of the thermal and turbulent energy contents of the IGrM on X-ray luminosity quantitatively. Moreover, a larger sample will help to obtain better statistics, including halos in the typical mass range of groups (i.e.\ $M\sim 10^{13} M_\odot$). This can be achieved by running nested-grid simulations of a number of filaments in different regions. To detect breaks in scaling relations under comparable conditions, many halos in the mass range of clusters have to be computed with a resolution that is sufficiently high for the sensible application of star formation and feedback recipes. Although we were not able to extend our analysis to objects outside of the zoom-in region because of too coarse resolution, our simulations point at a break down of self-similarity with respect to halo mass, while second self-similarity prevails in group and galaxy halos. 

\begin{acknowledgements}

We thank Surajit Paul and Luigi Iapichino for discussions that initiated the work presented this paper. Moreover, comments by Brian O'Shea helped us to improve our manuscript. PG acknowledges funding by NASA Astrophysics Theory Program grant \#NNX15AP39G. The simulations presented in this article were performed on SuperMUC(-NG) at the Leibniz Supercomputing Centre (project pr62ze). We also acknowledge the yt toolkit by \citet{Turk2011} that was used for our analysis of numerical data.

\end{acknowledgements}

\bibliographystyle{aa} % style aa.bst
\bibliography{references}

%\begin{thebibliography}{}

%\end{thebibliography}

%\section{Constraining the CGM/IGrM}
%\label{sec:constr}

\end{document}